\RequirePackage{fix-cm}
\documentclass[smallcondensed]{svjour3}     
\smartqed
\usepackage{graphicx}
\usepackage{bm}

\usepackage{amsthm}

\usepackage{mathtools}
\usepackage{amsfonts}

\newcommand\numberthis{\addtocounter{equation}{1}\tag{\theequation}}
\title{Enabling Imitation-Based Cooperation in Dynamic Social Networks
}

\author{Jacques Bara \and Paolo Turrini\footnote{Corresponding author.} \and Giulia Andrighetto}

\authorrunning{Bara, Turrini and Andrighetto} 

\institute{           
            Department of Mathematics \at
                University of Warwick, United Kingdom \\
              \email{jack.bara@warwick.ac.uk}
           \and
            Department of Computer Science \at
                University of Warwick, United Kingdom \\
              \email{p.turrini@warwick.ac.uk}
           \and
              Institute of Cognitive Sciences and Technologies \at
                National Research Council, Italy \\
              \email{giulia.andrighetto@istc.cnr.it}
}

\begin{document}
\maketitle

\begin{abstract}
The emergence of cooperation among self-interested agents has been a key concern of the multi-agent systems community for decades. With the increased importance of network-mediated interaction, researchers have shifted the attention on the impact of social networks and their dynamics in promoting or hindering cooperation, drawing various context-dependent conclusions. For example, some lines of research, theoretical and experimental, suggest the existence of a threshold effect in the ratio of timescales of network evolution, after which cooperation will emerge, whereas other lines dispute this, suggesting instead a Goldilocks zone. In this paper we provide an evolutionary game theory framework to understand coevolutionary processes from a bottom up perspective - in particular the emergence of a cooperator-core and defector-periphery - clarifying the impact of partner selection and imitation strategies in promoting cooperative behaviour, without assuming underlying communication or reputation mechanisms. In doing so we provide a unifying framework to study imitation-based cooperation in dynamic social networks and show that disputes in the literature can in fact coexist in so far as the results stem from different equally valid assumptions. 
\keywords{Social Networks \and Evolutionary Game Theory \and Partner Selection \and Imitation \and Emergence of Cooperation }
\end{abstract}

\section{Introduction}
From social media, to power grid \cite{powergrid} and road systems \cite{roads}, to mycorrhizae connecting trees \cite{mycorrhizal}, networks play a major role in promoting desirable behaviour. They can improve transport of nutrients or goods \cite{nutrient-transport}, they can connect long lost friends \cite{Watts1998,small-world-review} and they can even make computer systems resilient to attack \cite{HUTCHISON201813}. 

The multi-agent systems community has been greatly concerned with social networks and their effect on interaction since its very start. Social network analysis has been used to derive and extract systems of reputation \cite{Sabater2002,Pujol2002}, to explain the emergence of cooperation \cite{Gilbert1995,Salazar2011} or conventions \cite{Airiau2014} and to explore mechanisms of ostracism \cite{PerreaudePinninck2010}. Networks have also been used as a policy/control tool to prevent polarisation \cite{Santos2021}, to promote cooperation via partner selection \cite{Santos2019,Santos2020-sc} while also being an emergent property out of reciprocity \cite{Phelps2013}.


When the network is formed by self-interested agents engaging in strategic interaction, some desirable global properties, i.e., achieving Pareto-optimal outcomes, can be hindered by the desire to achieve higher payoff for themselves.
This is particularly true in games such as the Prisoner's Dilemma, when the benefits of ``cooperation" are outweighed by those of ``defection", but the latter is the individually rational choice.

Cooperation may mean contributing to some collective-risk social dilemma (e.g. investments in green energy) to avoid a collective catastrophe to occur, while defection means no contribution at all \cite{Smirnov2019}; in other settings cooperation may mean incurring some kind of cost (e.g. a tax) to receive social services and defecting means tax-avoidance while benefiting from said services \cite{GOTTLIEB198581}. 

How do social networks, where agents connect and interact with one another, affect their levels of cooperation? And what happens when these connections change in time?   Some theoretical results \cite{Gracia-Lazaro2012a} backed by empirical evidence \cite{Gracia-Lazaro2012b} have suggested that networks have little effect on cooperation or contribution \cite{Suri-Watts-2011}. However, these findings have been obtained using static networks, while a positive effect on cooperation was found experimentally in dynamic \cite{Rand2011,Wang2012} and temporal \cite{Li2020} networks with some analytic backing \cite{Santos-et-al-2006,Pacheco2006,Pinheiro2016}. Even static networks were found to improve cooperation in experiments \cite{Rand2014}, simulations \cite{Santos2005} and theory \cite{Ohtsuki2006,Pinheiro2016,Fotouhi2019}, though it seems to be the case that dynamic networks are far more amiable to cooperation than their static counterparts \cite{Melamed2017}. 


Despite these discrepancies, experimental results have broadly agreed that those who cooperate tend to be more popular than their misbehaving counters. When networks are dynamic and ties are at least partly endogenous, subjects rarely broke links with a cooperator \cite{Rand2011} partially causing them to have higher degree \cite{BRAVO2012481}. In fact as cooperators attract preferential attachment \cite{Santos2005,Poncela2008,Pinheiro2016}, they emerge as ``leaders" with high payoff \cite{Eguiluz2005}. Assortativity for mutually cooperative links arises out of subjects avoiding defectors when connections are formed bilaterally \cite{Wang2012} and via unfriending misbehaving neighbours formed unilaterally \cite{Fehl2011,Rand2011}.

Through ostracism \cite{Masclet2003} and punishment (for example sanctions as a form of costly ostracism \cite{PerreaudePinninck2010}), the co-evolutionary process generates networks with scale-free degree distributions, that promote even more cooperation \cite{Santos2005}, and are heavily clustered \cite{Fehl2011} around cooperators \cite{REZAEI2012}. Imagine, thus, a highly-interconnected sub-graph or core of cooperators, collectively working and benefiting one another, surrounded by parasitic defectors in the periphery, mostly avoiding one another while clinging to cooperators in the core. 

There is now solid theoretical and empirical evidence that reputation is an important mechanism for promoting cooperation \cite{nowak1998,milinksi2002,cuesta2014,pfeiffer2012,Sabater2002}. Here we take an alternative approach and look at the emergence of cooperation when reputation or other communication mechanisms are not available or not reliable enough, and simply focus on the effects of imitation strategies and network dynamics on promoting cooperation. We see our results as complementing reputation research, by showing, among others, when cooperation cannot be sustained by imitation and partner selection alone.


One method to completely decouple reputation-like mechanisms from the topological process is to impose partnerships exogenously. In doing so the importance of time scales is highlighted; when edge activity is ``bursty" - i.e. narrow, sudden spikes of activity -  cooperation is impeded while intermediate temporality analytically maximises cooperation \cite{Li2020}. Although other studies found a similar Goldilocks zone - maximal gain for an intermediate value(s) of input(s) - in the time scales \cite{Shirado2013}, much of the literature discusses a threshold effect or time scale separation. Some experiments have reported no evidence for such a threshold \cite{Wang2012} while others point to its existence \cite{Rand2011,Jordan2013}, backed by a slew of theoretic work \cite{Pinheiro2016,Pacheco2006,Santos-et-al-2006} that have direct analogues in percolation theory \cite{Parshani_2010,Deprez2015}. The lack of evidence seen by \cite{Wang2012} comes from considering too restricted a span of time scales, specifically the ratio between strategy and tie update rates, and thus requires further experiments to verify.

Such conflicted literature reflects the wide variety of often disjoint assumptions, frameworks and foundational concepts, at least in terms of temporal aspects. There is certainly a disconnect between theory and experiment; the former mostly considers large time evolution \cite{Pinheiro2016} or steady-state results \cite{Pacheco2006} while experiments are typically limited to short term studies on the scale of tens of rounds \cite{Rand2011,Rand2014,Wang2012}. There is then great variety in partner selection (e.g. random, round-robin, preferential) for experiments or dynamic tie updates for simulations, occurring in different regimes of the ratio between strategic and topological time scales \cite{Santos-et-al-2006,Pacheco2006,Fu2008}. Moreover, such experiments rarely, if ever, vary the ratio of timescales and as such there is very little empirical data regarding timescale separation. Understanding systematically why theories offer very different predictions, each backed by their own set of experiments, is a crucial task to truly understand cooperation. The core of such a task is to identify where fundamental assumptions align, differ or are compatible in a systematic way that has yet to be undertaken by current research.


\subsection{Contribution}
\begin{figure}[thp]
\centering
\includegraphics[width=119mm]{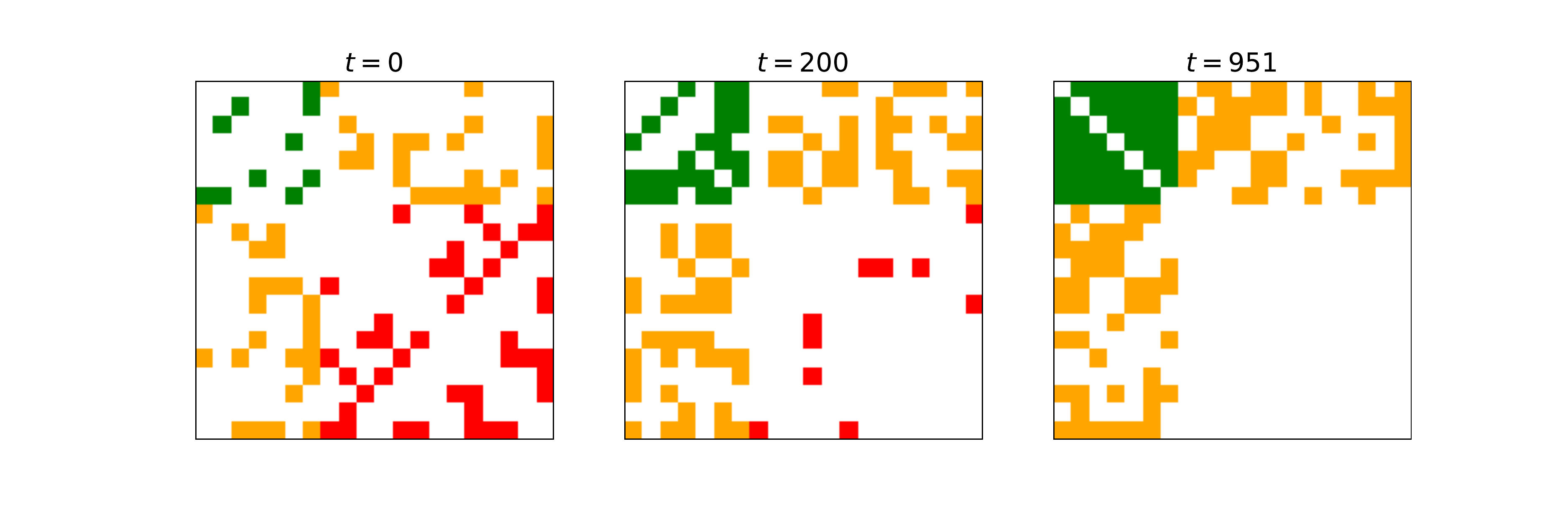}
\caption{The evolution of a dynamic network following the extreme popularity update model, wherein cooperators (C) are always befriended, defectors (D) are always unfriended and strategies are fixed. Colours represent the type of edge: CC edges are in green, CD edges in orange and DD edges in red. Agents are initially placed on a random graph generated by the Erd\"{o}s-R\`{e}nyi model with probability $p=0.2$ then allowed to update every timestep. As $t\to\infty$ a core-periphery structure emerges where cooperators inhabit the core and defectors loosely hang on in the periphery. Note that so long as a non-zero number of cooperators $C>0$ remain any initial condition, under any of the partner-update rules, will eventually stabilise to this type of core-periphery configuration with a core of size $C$.}
\label{fig:core-periphery}
\end{figure}

We propose an evolutionary game-theory framework, which call the Cooperative And Networked DYnamic (CANDY) framework, to disentangle the basic assumptions that enable cooperation in dynamic social networks, only relying on basic reality-resembling imitation strategies. CANDY starts with assumptions on agents' decision-making - how someone decides to cooperate/defect (the imitation strategy) and chooses to befriend/unfriend others (the network evolution) - and produces the resulting average payoff and total number of cooperators. An illustration of the type of results CANDY can produce is given in Figure \ref{fig:core-periphery}. 

Methodologically, this framework allows for rigorously testing and comparing different assumptions, finding sets of assumptions which are compatible with empirical data and overcome much of the literature heterogeneity. For instance one may suspect the division in whether the ratio of timescales produces a threshold or Goldilocks effect to be due to the differences in assumptions theorists made or in exogenous update rules imposed upon test subjects. Our framework allows the coexistence of such incompatible results by rigorously scrutinising the underlying assumptions.

To illustrate the power of the CANDY framework, we recover the theoretical results \cite{Pacheco2006} for when edges undergo a birth-death process and strategies follow a Moran \cite{Nowak2004,TAYLOR2004} or Wright-Fisher \cite{Imhof2006} process. Furthermore, we consider other edge update models (such as cooperator-popularity) and behavioural models (such as conditional cooperation) that capture the assumptions of other research lines, in order to illustrate the qualitative differences that emerge. 

Moreover, we provide a nuanced discussion on timescale separation by reproducing both the threshold effect as seen in \cite{Rand2011,Jordan2013,Pinheiro2016,Pacheco2006,Santos-et-al-2006} and a Goldilocks zone for defection; in many cases we suspect such effects are really artefacts of the finite number of rounds occurring. In recovering both phenomena we highlight how sensitive results are to both initial conditions and to the assumptions of the researcher. That is to say, by having slightly different but equally valid assumptions, the qualitative results can be significantly different.

Finally we illustrate how one assumption/observation - that cooperators are more popular - can lead to the emergence of the same core-periphery structure, despite instrumentally different update rules. Such core-periphery structures have been observed in experiments with human subjects \cite{Sohn2019} and other agent-based models \cite{Shepherd2020,REZAEI2012}, even when edges can only be broken, not formed for agents with multidimensional opinion spaces; there is an intuitive correspondence between our cooperators and defectors to homogeneous and adversarial agents respectively seen in \cite{Shepherd2020}. 

\subsection{Paper Structure}

Section \ref{sec:theory} presents the mathematical setup, followed by the introduction of the CANDY framework. Section \ref{sec:partner} analyses the partner selection mechanics, while Section \ref{sec:strategy} looks at the imitation strategies. Section \ref{sec:results} provides the main results, showing the behaviour of the CANDY framework on key (random) graph models. We then move to the discussion of the findings and some key pointers for future research. In appendix we provide some  basic preliminaries on variables, expectations and dynamical systems.

\section{Theoretical Model}\label{sec:theory} 


For a simple graph $G=(V,E)$ of $N$ nodes playing a repeated Prisoner's Dilemma game, denote the adjacency matrix as $A = (a_{ij}:i,j\in V)$ and the strategies as a vector $\bm{s}=(s_i:i\in V)$, where the binary strategy $s$ can either be $1$ (cooperate) or $0$ (defect)\footnote{The strategies behave as indicator functions to cooperate. In this way we can build further indicator functions for more complicated situations involving two people $i,j$; for example $\bm{1}($both cooperate$)=s_is_j$ and $\bm{1}($both defect$)=(1-s_i)(1-s_j)$.}. A cooperators pays a cost $c$ per neighbour, such that each of her neighbours gains a benefit $b$. A defector, on the other hand, pays nothing and nothing happens. This payoff structure was chosen to match the predominant games considered in the literature.

From these $N(N+1)$ local variables we can find global/aggregate variables that are of the most interest. Given $A$ and $\bm{s}$ we can find the payoff vector $\bm{\pi} = (\pi_i)$ by defining a modified Laplacian using the payoff-structure $(b,c)$
\begin{align}
    L' = cK - bA \label{eq:modified_laplacian}
\end{align}

where $K=diag(k_i)$ is the diagonal matrix of degrees. Note we will also be using $\bm{k} = (k_i)$ to refer to the vector of degrees. The payoff vector is then related to the strategy vector by a simple transformation
\begin{equation}
    \bm{\pi} = -L'\bm{s} \label{eq:payoff_vector}
\end{equation}

of which the average payoff $\bar{\pi}$, can be simply found.
\begin{equation}
    \bar{\pi} = \frac{b-c}{N}\bm{k}\cdot \bm{s}
\end{equation}

As with most of the evolutionary game theory literature the two main quantities of concern are the fraction or number of cooperators and the average payoff (i.e. payoff per capita). The latter is less often considered, despite the insight it provides into the actual network structure and thus helps to understand both processes occurring. For reference, therefore, we write below the equations for the total number of cooperators $C$ and the average payoff $\bar{\pi}$ in terms of local variables.
\begin{gather}
    C = \sum_{i\in V} s_i \label{eq:C}\\
    \bar{\pi} = \frac{b-c}{N}\sum_{i,j\in V}a_{ij}s_j \label{eq:pibar}
\end{gather}

Rather than dealing with strictly discrete variables, we can instead move to the continuum by considering probabilities - here such probabilities are also identically the expectations of binary variables. Specifically we will consider $\tilde{a}_{ij}\equiv \mathbb{P}(a_{ij}=1)$ and $\tilde{s}_i \equiv \mathbb{P}(s_i=1)$ as the probabilities for edge $(i,j)$ to exist and for node $i$ to cooperate at time $t$, respectively. We then assume such probabilities evolve due to two independent processes - a vector field acting on the adjacency matrix, $f_g$, and one acting on the strategy vector $f_s$. In other words in the joint probability space $[0,1]^{N^2}\times [0,1]^N$ where a point represents an entire state, this point moves due to the 'velocities' $f_g$ and $f_s$.

\subsection{CANDY Framework}

In general we can write down $N(N+1)$ coupled differential equations (DEs) for both success probabilities in terms of our two vector fields $f^g = (f^g_{ij}:i,j\in V)$ and strategic process $f^s=(f^s_i:i\in V)$. 
\begin{gather}
    \frac{d\tilde{a}_{ij}}{dt} = f_{ij}^g(A,\bm{s},t;\tau_g) \label{eq:a_ijdot}\\
    \frac{d\tilde{s}_{i}}{dt} = f_i^s(A,\bm{s},t;\tau_s) \label{eq:s_idot}
\end{gather}

More often than not it is easier to construct transition rates rather than the full differential equation. Thus let us write $f^g_{ij}$ and $f^s_{i}$ in terms of transition rates: given $(i,j)\not\in E$ the rate to form said edge $g_{ij}(0,1)$; given $(i,j)\in E$ the rate to break the edge $g_{ij}(1,0)$; given $s_i=0$ the rate to cooperate $h_i(0,1)$ and finally given $s_i=1$ the rate to defect $h_i(1,0)$\footnote{Strictly speaking such transition rates depend on the realisation/observation/sample of the Bernoulli variables, \textit{not} on the respective parameters - in other words dependence is on the non-tilde quantities. However where necessary we approximate $f^g$ and $f^s$ by substituting the explicit dependence on realisation with the expected values (i.e. the success probabilities).}.

\begin{gather}
    f^g_{ij} = g_{ij}(0,1)(1-\tilde{a}_{ij}) - g_{ij}(1,0)\tilde{a}_{ij} \label{eq:f_g-general} \\
    f^s_i = h_i(0,1)(1-\tilde{s}_i) - h_i(1,0)\tilde{s}_i \label{eq:f_s-general}
\end{gather}

This provides the basis for the entire \textbf{C}ooperative \textbf{A}nd \textbf{N}etworked \textbf{DY}namics (CANDY) framework. So long as the update rules have a closed form, the above $N(N+1)$ equations fully specify the dynamics. CANDY allows for update rules that are time-dependent, parameterised and/or heterogeneous - suffice to say an incredibly broad range of possibilities. Moreover, by integrating the vector fields $f^g$ and $f^s$, the flows are fully recovered $\Phi = (\Phi^g,\Phi^s)$ allowing researchers to sidestep lengthy and computationally heavy agent-based simulations.

The evolution of $C$ and $\bar{\pi}$ are further gotten by differentiating Eqs. \ref{eq:C} and \ref{eq:pibar} to get DEs in terms of the generalised processes.

\begin{align}
    \frac{dC}{dt} &\equiv F^C = \sum_{i\in V} f^s_i \label{eq:dcdt}\\
    \frac{d\bar{\pi}}{dt} &\equiv F^\pi = \frac{b-c}{N}\sum_{i,j\in V} (f^g_{ij}\tilde{s}_i + \tilde{a}_{ij}f^s_i) \label{eq:dpidt}
\end{align}

In specifying the local evolution due to $f^g$ and $f^s$, researchers are able to clearly lay out their assumptions, numerically integrate and finally compare predictions. This generative method follows the same ideology as much of the agent-based modelling community, but with the added bonus of allowing for comparisons between qualitatively different hypotheses as it provides a standard framework to work within. 

In general solving for the two global variables requires local knowledge and solutions may not be analytically tractable. However under certain local processes, $F^C$ and $F^\pi$ may be written only in terms of $C$ and $\bar{\pi}$ (and generally $t$), in which case the global behaviour reduces significantly down to a system of 2 coupled DEs, or even a single equation. When analytic solutions do not exist, numerical integration still provides a marked computational improvement on pure agent-based simulations. 

In the next two sections we look at the two dimensions of the coevolutionary process, that of partner-updates and that of strategy-updates. We treat the two processes as independent, much like the equations of a fluid flow can be broken down into component parts.

\section{Partner-Update Rules}\label{sec:partner}
In this section we consider a broad range of graph-theoretic models (GMs) of partner updates, in other words how edges change over time. We do so to illustrate the variety of empirical observations and assumptions of relationship building/breaking that are as valid as one another but that may produce entirely different dynamics. In particular we highlight three GMs - an extreme version of empirical observations that cooperators are always popular, an active linking model \cite{Pacheco2006} and exogenously imposed networks as in \cite{Li2020}.

In later sections we will combine such models with different behavioural models (see the next section) of how agents choose to cooperate or defect. In so doing we arrive at a diverse set of dynamics, some producing timescale separation while others showing regimes of mass cooperation and mass defection.

\subsection{Graph-theoretic Model 1: Extreme Popularity}
Following empirical results that cooperators are more popular \cite{Rand2011,Wang2012,BRAVO2012481}, we take the extreme limit where only cooperators are befriended and defectors are entirely unfriended. That is at each time step (discrete round) a pair of nodes $i,j\neq i \in V$ is randomly chosen. If the alter $j$ is a cooperator, $s_j=1$, then the ego $i$ will unilaterally form an edge with $j$ if none previously existed. Otherwise if $j$ is a defector, $s_j=0$, $i$ will unilaterally break ties with $j$ if the edge already existed. Note that by specifying \textit{at each time step} we have inadvertently set a graph-theoretic timescale $\tau_g$; in the units of time-steps $\tau_g=1$ but one could also use units defined by the total number of edges possible $\tau_g=N(N-1)/2$. 

From this, we can specify the transition rate for an edge $(i,j)$ to form $g_{ij}(0,1)$ or break $g_{ij}(1,0)$. 
\begin{align}
    g_{ij}(0,1) &= \frac{1}{\tau_g} \frac{1}{2} \bigg[s_i(1 + s_j) + (1-s_i)s_j\bigg] \label{eq:f_g-01} \\
    &= \frac{1}{\tau_g} \frac{s_i+s_j}{2}\\
    g_{ij}(1,0) &= \frac{1}{\tau_g} \frac{1}{2} \bigg[s_i(1-s_j) + (1-s_i)\big[2(1-s_j) + s_j\big] 
    \bigg] \label{eq:f_g-10} \\
    &= \frac{1}{\tau_g}\big( 1 - \frac{s_i + s_j}{2}\big)
\end{align}

The factor of $1/2$ is the probability to choose the active node as the ego, i.e. the one to unilaterally make or break a tie. The first term in Eq. \ref{eq:f_g-01} is from if $i$ is a cooperator and $j$ is chosen as ego, $j$ always connects, otherwise if $i$ is chosen as ego then $i$ connects only to another cooperator; the second term is when $i$ is a defector and is the ego, thus can connect to $j$ if $j$ is a cooperator. Similarly, for Eq. \ref{eq:f_g-10}, the first term comes from when $i$ is a cooperative ego disconnecting to a defector; the latter half is when $i$ is a defector so either if both $i,j$ are defectors there's two ways to break the tie or a cooperator $j$ disconnects to $i$. Finally the evolution of $\tilde{a}_{ij}$ can be found by summing the gain terms and loss terms.
\begin{align}
    f^g_{ij}&= \frac{1}{2\tau_g}\bigg[ (s_i+s_j)(1-\tilde{a}_{ij}) - (2 - s_i-s_j)\tilde{a}_{ij}\bigg] \nonumber \\
    &= \frac{1}{\tau_g} \bigg[ \frac{s_i+s_j}{2} - \tilde{a}_{ij}\bigg] \label{eq:f_g-extreme-popularity} 
\end{align}

As a sanity check when both $i$ and $j$ cooperate and are connected, $f^g = 0$ so the edge is stable. Similarly when both are defectors and disconnected $f^g=0$ so the lack of an edge is stable too. Finally in general, as strategies are not fixed, to compensate for their probabilistic nature, we can simply replace $s_i$ with $\tilde{s}_i$. The utility of having such a kernel is that we can obtain an ODE for $k_i$ - simply sum over all $j\neq i$, $\dot{k}_i = \sum_{j\neq i}f^g_{ij}$ - and by extension an ODE for the payoff per capita when strategies are fixed.


\subsection{Graph-theoretic Model 2: Active Linking}
We consider the active linking model \cite{Pacheco2006} to illustrate how their assumptions lead to a bottom-up or micro-scale model that then lead to the same meso-scale and macro-scale results. In active linking, cooperators and defectors form edges at some constant rate $\alpha_C$ and $\alpha_B$ and that different edge types decay away at rates $\beta_{CC},\beta_{CD}$ and $\beta_{DD}$ for $CC, CD$ and $DD$ edges respectively. Notice unlike our first model, this assumes an agent is not biased towards befriending cooperators over defectors, counter to empirical results. Instead your strategy impacts how strongly you connect to anyone else; in a sense this views relationships as stemming from the individual and not from reputations/biases.

Similarly to extreme popularity we can arrive at the transition rates by considering when $i$ is a cooperator then $j$ is either cooperator or defector and vice-versa. Notice now instead of a single graph-theoretic timescale there are potentially several timescales (edge formation versus edge destruction), but where they are not significantly different one could simply take an average.
\begin{align*}
    & \begin{aligned} \mathllap{g_{ij}(0,1)}=
    s_is_j\alpha_C^2 + (1-s_i)(1-s_j)\alpha_D^2 \\+ (s_i+s_j-2s_is_j)\alpha_C\alpha_D  \end{aligned}\\
    & \begin{aligned} \mathllap{g_{ij}(1,0)}= s_is_j\beta_{CC} + (1-s_i)(1-s_j)\beta_{DD} \\+ (s_i+s_j-2s_is_j)\beta_{CD} 
    \end{aligned}
\end{align*}

Substituting the above transition rates into Eq. \ref{eq:f_g-general} we get an incredibly long and verbose equation for $f^g_{ij}$.
\begin{multline}
    f^g_{ij} = s_is_j\big[\alpha_C^2(1-\tilde{a}_{ij})- \beta_{CC}\tilde{a}_{ij}\big] \\ 
    + (s_i+s_j-2s_is_j)\big[\alpha_C\alpha_D(1-\tilde{a}_{ij}) -  \beta_{CD}\tilde{a}_{ij}\big] \\
    + (1-s_i)(1-s_j)\big[\alpha_D^2(1 - \tilde{a}_{ij})- \beta_{DD}\tilde{a}_{ij}\big] \label{eq:f_g-active-linking}
\end{multline}

Importantly, each of the three terms above correspond to the three types of possible edges forming and breaking. The first term ($s_is_j[\cdots]$) denotes evolution of $CC$-edges since $s_is_j=1$ iff both $i$ and $j$ are cooperators; similarly the other two denote evolution of the $CD$ and $DD$-edges respectively. 

\subsection{Graph-theoretic Model 3: Exogenously Imposed Networks}
Consider when the network is exogenously imposed - perhaps the structure is synthetically produced or reshuffled by the researcher, or as in \cite{Li2020} the network is empirical, such as contact-networks. In this case $N^2$ equations are eliminated and the only interesting dynamics occurs within the strategic space, as the graph-theoretic space has been fully specified. 

For instance in the discrete case, let $\{t_0,t_1,\cdots\} \in \mathbb{R}$ be a sequence of timestamps at which the adjacency matrix has changed so that during each interval $t_m\leq t < t_{m+1}$ the network is fixed. The graph-theoretic timescale can be defined as a statistic on the set of interval periods, for example the mean or the minimum of such a set. When strategies update following an imitate-payoff type rule (see BM1 below) slower than the network gets updated, cooperation is promoted while when $\tau_s<\tau_g$ mass defection occurs \cite{Li2020}.

\section{Strategy-Update Rules}\label{sec:strategy}
In this section we focus on how agents decide to cooperate or defect, in particular we consider three different mechanisms for the imitation of neighbours. One, when imitation is purely payoff-dependent of a random successful neighbour, similar to \cite{Santos2005}. Two, imitation happens due to social pressures that is presented with a crowd of opinions disagreeing with her, an agent will change her strategy, in other words conditional cooperation \cite{FISCHBACHER2001397,Burton-Chellew1291} which is also known as the voter model \cite{Holley1975}. Finally, as a hybrid of the previous two, where both payoff and the will of the crowd matters, the pairwise comparison rule \cite{Pinheiro2016}. 

There are, of course, a multitude of other more complex decision making processes - for example moody conditional cooperation \cite{Grujic2014} and tag-based cooperation when agents have observable traits \cite{Stivala2016} - however to address every single possible rule would be never-ending. Hence for the sake of a focused scope and for readability we focus only on three.

\subsection{Behavioural Model 1: Imitate-Payoff}
Here we consider strategies are updated by pure imitation that are discretely payoff-dependent, that is imitation occurs only iff the proposed alter has a higher payoff. Every $\zeta$ time-steps, $\eta$ existing \textit{edges} are picked randomly\footnote{Strictly speaking, $\min(|E|,\eta)$ edges are chosen in the event there are fewer than $\eta$ edges that actually exist.}. One node per edge is then chosen randomly to be the ego $i$, who will imitate the strategy of their partner $j$ iff $\pi_j > \pi_i$. Measuring in time-steps the strategic timescale is $\tau_s = \zeta/\eta$, while relative to the timescale due to the graph-theoretic process $\tau_s= \tau_g(\zeta/\eta)$.

\subsection{Behavioural Model 2: Conditional Cooperation}
Alternatively to the payoff-dependent model, we can instead consider conditional cooperation (also known as a voter model) \cite{Burton-Chellew1291,Holley1975} where an agent cooperates if their neighbours also cooperate. This is frequency-dependent and payoff-independent, embodying the social pressures to imitate and change strategies (see for example \cite{Bara2021,Yildiz2010,Vazquez_2008}). Every $\zeta$ timesteps, $\eta$ \textit{nodes} are picked randomly. Such a node $i$, with strategy $\sigma$, will switch strategies to $\sigma'$ iff the fraction of their neighbourhood with strategy $\sigma'$ strictly exceeds a given threshold $v_{\sigma\sigma'}$. This rule is thus parameterised by two thresholds $\mathbf{v}=(v_{cd},v_{dc})$, where $v_{cd}$ is the for a cooperator to start defecting - that is the minimum fraction of a cooperator's neighbourhood that are defectors - and similarly for $v_{dc}$. Since nodes are picked, rather than edges, the most reasonable time unit to consider would be in timesteps hence $\tau_s=\zeta/\eta$, otherwise when compared to the graph-theoretic process $\tau_s=\frac{N-1}{2}\frac{\zeta}{\eta}\tau_g$.

\subsubsection*{Probabilistic Version}
Rather than having deterministic rules, we let the $\eta$ nodes update stochastically, depending on the fraction of cooperators (or defectors) in their local neighbourhood. In this treatment, the probability for a defector to cooperate is the fraction of their neighbour who cooperate and vice versa.
\begin{align*}
    h_i(0,1) &= \frac{1}{\tau_s}\frac{\sum_{j}a_{ij}s_j}{\sum_{j}a_{ij}} \\
    h_i(1,0) &= \frac{1}{\tau_s}\big(1-\frac{\sum_{j}a_{ij}s_j}{\sum_{j}a_{ij}}\big) \\
    f^s_i &= \frac{1}{\tau_s}\bigg(\frac{\sum_{j}a_{ij}s_j}{\sum_{j}a_{ij}} - \tilde{s}_i\bigg) \numberthis \label{eq:f_s-ImF}
\end{align*}

Unfortunately due to the denominator, taking the expectation of Eq. \ref{eq:f_s-ImF} does not result in substituting the variables for their probabilities. Consider the first term as a function $\sigma_i:\mathcal{A}\times\mathcal{S}\rightarrow\mathbb{R}$ acting on the adjacency matrix and the strategy vector. Denote, as usual, the expected values as $\tilde{A}=(\tilde{a}_{ij}:i,j\in V)$ and $\tilde{\bm{s}}=(\tilde{s}_{i}:i\in V)$. We can approximate $\sigma_i$ with a Taylor expansion.
\begin{multline*}
    \sigma_i(A,\bm{s}) = \sigma_i(\tilde{A},\tilde{\bm{s}}) + \sum_j \bigg[ (a_{ij}-\tilde{a}_{ij})\frac{\partial \sigma_i}{\partial a_{ij}} + (s_j-\tilde{s}_j)\frac{\partial \sigma_i}{\partial s_j} \bigg] \\
    + \frac{1}{2}\sum_j\bigg[ (a_{ij}-\tilde{a}_{ij})^2\frac{\partial^2 \sigma_i}{\partial a_{ij}^2} + (s_j-\tilde{s}_j)^2\frac{\partial^2 \sigma_i}{\partial s_j^2} \bigg] \\
    + \sum_{j,k}\bigg[(a_{ij}-\tilde{a}_{ij})(a_{ik}-\tilde{a}_{ik})\frac{\partial^2 \sigma_i}{\partial a_{ij}a_{ik}} \\+ (s_j-\tilde{s}_{j})(s_{k}-\tilde{s}_{k})\frac{\partial^2 \sigma_i}{\partial s_{j}s_{k}} 
    + (a_{ij}-\tilde{a}_{ij})(s_{k}-\tilde{s}_{k})\frac{\partial^2 \sigma_i}{\partial a_{ij}s_{k}} \bigg] +\cdots 
\end{multline*}

Taking the expectation of the Taylor expansion, the first order derivative terms drop away while the higher order terms are now multiplied by variances and covariances; the variance can be explicitly found in terms of the Bernoulli parameters $Var(a_{ij}) = \tilde{a}_{ij}(1-\tilde{a}_{ij})$. Any second order derivatives (or higher) with respect to strategies disappear as $f_i$ are linear in strategies. Finally assuming pointwise independence, covariance terms will vanish, thus leaving an approximate expression for $f_i$ in terms of only the expected adjacency matrix and strategy vector.
\begin{equation*}
    \mathbb{E}[\sigma_i(A,\bm{s})] \approx \frac{\sum_{j}\tilde{a}_{ij}\tilde{s}_j}{\sum_{j}\tilde{a}_{ij}} +
    \sum_j \frac{\tilde{a}_{ij}(1-\tilde{a}_{ij})}{(\sum_{k}\tilde{a}_{ik})^2} \bigg(\frac{\sum_{k}\tilde{a}_{ik}\tilde{s}_k}{\sum_{k}\tilde{a}_{ik}} - \tilde{a}_{ij}\bigg)
\end{equation*}
Noting that the second term and higher order terms are $O(N^{-2})$ for non sparse graphs, a first order approximation of $\sigma_i$ is sufficient so we can in fact replace the variables in Eq. \ref{eq:f_s-ImF} with the continuous version.

\subsection{Behavioural Model 3: Pairwise Comparison Rule}
In Model 1 (imitate-payoff) an agent imitates precisely one neighbour based on the pairwise difference in payoffs, following a step-function. In reality someone may copy a friend even if their payoff was worse; for a start there might be some leniency where a friend might earn \$10 more but I do not instantly side with him. Moreover, although a single friend may earn more via their alternative strategy, there may be social pressures to stick to my current strategy. Model 2 takes the extreme and assumes the only thing that matters is the number of friends with a given strategy, so that if $80\%$ of my neighbourhood are cooperators I'm $80\%$ likely to cooperate.

The birth-death process of \cite{Pinheiro2016} takes a hybrid/intermediate stance relative to Models 1 and 2. It assumes that a) payoffs matter for imitation but not as rigidly as a step function - instead it should be smooth and continuous - and b) that social pressures also matter, so that the probability to flip strategies is given by the average probability to imitate an alternative strategy. Explicitly, the probability $p_{ij}$ for $i$ to imitate the alternative strategy of $j$, i.e. $s_i \neq s_j$, follows a sigmoid with parameter $\beta$ measuring the strength of peer pressure and payoffs acting as the energies. The probability for $i$ to flip strategies is the mean of pairwise $p_{ij}$ over alternative-strategy neighbours. It is worth noting here that the instrumentalisation of this process would involve choosing a number of \textit{nodes} rather than edges to update as such there will not be any factors of $2$ to differentiate between ego or alter. 
\begin{align}
    p_{ij} &= \frac{1}{1+\exp{[-\beta(\pi_j - \pi_i)]}} \label{eq:pairwise_comparison} \\
    h_i(0,1) &= \frac{1}{\tau_s}\sum_{j\in V} \frac{p_{ij}}{k_i}\tilde{s}_j a_{ij} \nonumber \\
    h_i(1,0) &= \frac{1}{\tau_s}\sum_{j\in V} \frac{p_{ij}}{k_i}(1-\tilde{s}_j) a_{ij} \nonumber
\end{align}

Note that as payoffs in $p_{ij}$ can be simplified by Eq. \ref{eq:payoff_vector}, $\pi_i = -\sum_{k\in V} L'_{ik}s_k$, we can rewrite the exponent in terms of only strategies and adjacency $\pi_j - \pi_i = \sum_{k\in V}(bs_k - c)(a_{jk}-a_{ik})$. The evolution of the cooperation probability is thus given below.
\begin{align}
    f^s_i &= \frac{1}{\tau_s}\sum_{j\in V}\frac{p_{ij}}{k_i}a_{ij}\big[(1-\tilde{s}_i)\tilde{s}_j - \tilde{s}_{i}(1-\tilde{s}_{j}) \big] \nonumber \\
    &= \frac{1}{\tau_s k_i}\sum_{j\in V}p_{ij}a_{ij}\big[\tilde{s}_j - \tilde{s}_i \big] \nonumber \\
    &= \frac{1}{\tau_s k_i}\sum_{j\in V}  \frac{a_{ij}(\tilde{s}_j - \tilde{s}_i)}{1+\exp[-\beta\sum_{k\in V}(bs_k-c)(a_{jk} - a_{ik})]} \label{eq:f_s-I3}
\end{align}

Once the network is dynamic, simply replace $a_{ij}$ with $\tilde{a}_{ij}$ for when adjacency is probabilistic. 

\section{Results}\label{sec:results}
In this section we present several results and discussions, some theoretical and some computational, in order to highlight the advantages of the framework. We begin with the graph-theoretic models sans strategy - that is cooperators and defectors are fixed but they may change friendships - to illustrate how the network evolves and the structures that emerge. As both extreme popularity and active linking produce similar core-periphery mesostructures, we then combine extreme popularity - being the simplest model - with the three behavioural models: imitate-payoff, conditional cooperation and pairwise comparison. In doing so we identify cases timescales separate and cases where they do not.

\begin{figure}[bthp]
\centering
\includegraphics[width=119mm]{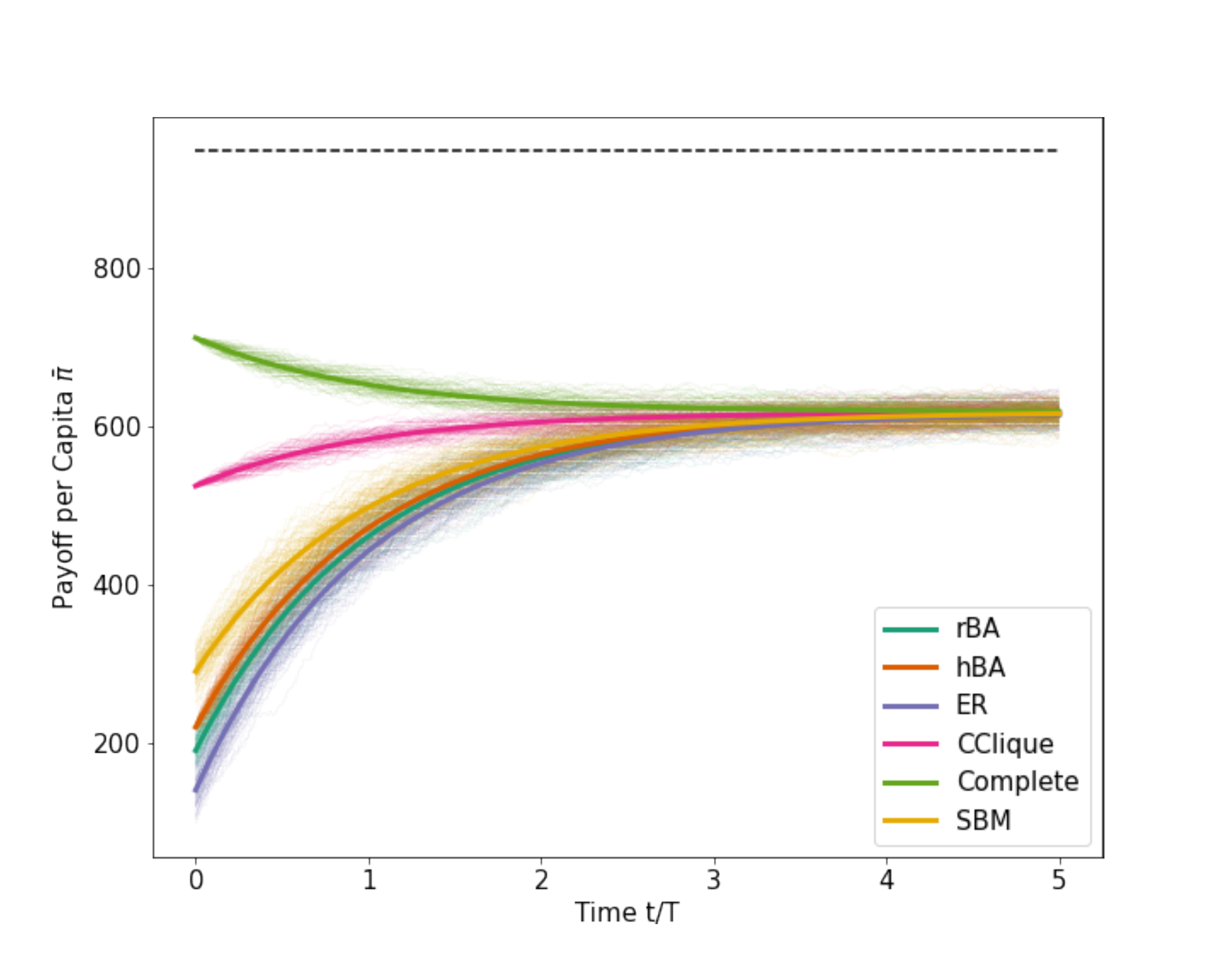}
\caption{Evolution of payoff per capita as agents play a Prisoner's Dilemma with fixed strategies and where cooperators have extreme popularity (see Graph-theoretic Model 1). Colours indicate the random graph model used to initialise a game (see Appendix for details); faint lines indicate individual runs while the bold lines represent the CANDY results, Eq. \ref{eq:pi_t_extreme_popularity}; the black dashed line is a baseline if all agents cooperated and were on a complete graph. Time has been rescaled by the number of dyads, $T=N(N-1)/2=190$, so that every $T$ time-steps on average all pairs have been updated once}
\label{fig:fixedstrategy}
\end{figure}

For our computational examples and illustrations in Figures \ref{fig:fixedstrategy}-\ref{fig:timescale_ratio_g1b3} we simulate using $N=20$ agents, 100 runs of a Prisoner's Dilemma with $(b,c)=(100,50)$ to be in line with the experiments of \cite{Rand2011}. Moreover, our simulations begin with a variety of initial network conditions, that is networks produced by different random graph generators.

\begin{itemize}
    \setlength\itemsep{1em}
    \item \textbf{Erd\"{o}s-R\`{e}nyi (ER)} - the canonical random graph model where pairs of nodes are connected with probability $p=0.2$.
    \item \textbf{Barab\'asi-Albert (BA)} - a preferential attachment model that grows a network of $N=20$ nodes by continuously adding nodes with $m=3$ edges.
    \begin{itemize}
        \item \textbf{Random assignment (rBA)} - the initial cooperators are randomly assigned.
        \item \textbf{Highest assignment (hBA)} - the initial cooperators are assigned to the nodes with highest degree.
    \end{itemize}
    \item \textbf{Cooperator Clique (CClique)} - the $C=15$ initial cooperators are completely connected to one another as a clique, while the remaining defectors are entirely disconnected from all others.
    \item \textbf{Complete} - all nodes are attached to one another.
    \item \textbf{Stochastic Block Model (SBM)} - specifically we use the assortative planted partition model \cite{SBM}, a special case of an SBM, with 2 communities, with an in-group edge probability of $p=0.8$ and out-group edge probability of $q=0.02$. 
\end{itemize}

\subsection{Fixed Strategy} 
By fixing strategies we can more clearly understand what graphs form as a consequence of $f^g$. In particular we see the emergence of mesoscale structures such as core-periphery (specifically the continuous core-periphery model \cite{BORGATTI2000}). The strength of such structures depend on how friendly and attractive cooperator/defectors are as friends as well as the perseverance of friendship types, however in most realistic settings (where cooperators are universally popular and defectors are avoided) a cooperator-core forms while a defector-periphery struggles to attach themselves to the core while largely avoiding one another. Henceforth for brevity we denote the mesoscale structure of cooperator-core-defector-periphery as CCDP.

\subsubsection*{Extreme Popularity}For the extreme popularity graph-theoretic model, Eq. \ref{eq:f_g-extreme-popularity} can be analytically integrated, thus leading to equations for $\tilde{a}_{ij}, k_i$ and $\bar{\pi}$.
\begin{gather}
    \tilde{a}_{ij}(t) = \bigg(\alpha_{ij} - \frac{s_i+s_j}{2}\bigg)e^{-t/\tau_g} + \frac{s_i+s_j}{2} \label{eq:aij_extreme_popularity} \\
    k_i(t) = \bigg(\kappa_{i} - \frac{Ns_i+C-2s_i}{2}\bigg)e^{-t/\tau_g} + \frac{Ns_i+C-2s_i}{2} \\
    \bar{\pi}(t) = \big(\bar{\pi}(0) - \bar{\pi}_* \big)e^{-t/\tau_g} + \bar{\pi}_* \label{eq:pi_t_extreme_popularity} \\
    \bar{\pi}_* = \frac{b-c}{2N}(N+C-2)C \label{eq:pistar_extreme_popularity}
\end{gather}

\begin{equation*}
    \bar{\pi}(t) = \big(\bar{\pi}(0) - \bar{\pi}_* \big)e^{-t/\tau_g} + \bar{\pi}_*
\end{equation*}

where $\alpha_{ij},\kappa_i$ and $\bar{\pi}(0)$ are the respective initial conditions. 

In order to supplement the theoretical prediction (Eq. \ref{eq:pi_t_extreme_popularity}) we run agent-based simulations following the extreme popularity model. As we see in Fig. \ref{fig:fixedstrategy} regardless of the initial condition - that is the network generation model used to initialise friendships - all converge to the same payoff per capita (Eq. \ref{eq:pistar_extreme_popularity}). Moreover, only a small subset of possible graphs can generate this value $\bar{\pi}_*$; for $X,Y$ the number of CC and CD edges respectively, $4X+2Y = (N+C-2)C$. 

In fact the graphs that form all exhibit core-periphery structure similar to what illustrated in Fig. \ref{fig:core-periphery}, regardless of initial condition and partner-update rule, so long as they follow the assumption that "cooperators are popular" and that there are $C>0$ cooperators - as we will see in the next subsection. In other words this cooperator-core defector-periphery (CCDP) is a stable configuration, along with an entirely disconnected networks of pure defectors. 


\subsubsection*{Active Linking} In the notation of the original paper \cite{Pacheco2006} denote the number of $CC,CD$ and $DD$-edges as $X,Y$ and $Z$ respectively; in terms of elements of the adjacency matrix, the edge-set sizes are simply the double sum over $i,j>i$ of the elements $a_{ij}$ multiplied by the relevant indicator function. In other words by taking the sum of each of the three terms in Eq. \ref{eq:f_g-active-linking} we recover exactly the evolutionary equations for $X,Y$ and $Z$ laid out originally \cite{Pacheco2006},
\begin{align*}
    \frac{dX}{dt} &= \alpha_C^2(X_m - X) - \beta_{CC}X \\
    \frac{dY}{dt} &= \alpha_C\alpha_D(Y_m - Y) - \beta_{CD}Y \\
    \frac{dZ}{dt} &= \alpha_D^2(Z_m - Z) - \beta_{DD}Z
\end{align*}
where $X_m, Y_m$ and $Z_m$ are the maximum sizes of the edge sets given a number of cooperators $C$ in the population. The steady state solution is thus given by
\begin{gather*}
    X_* = \frac{\alpha^2_C X_m}{\alpha^2_C + \beta_{CC}} \\
    Y_* = \frac{\alpha_C\alpha_D Y_m}{\alpha_C\alpha_D + \beta_{CD}} \\
    Z_* = \frac{\alpha^2_D Z_m}{\alpha^2_D + \beta_{DD}}
\end{gather*}
 
In a realistic setting cooperators rarely broke from each other \cite{Rand2011} so that $\beta_{CC}\approx0$ while defectors are regularly unfriended by everyone \cite{Fehl2011} $\beta_{DD}\gg \alpha_D^2$. We see therefore that $X_*\approx X_m$, $Z_*\ll Z_m$ and $Y_*\in [0,Y_m]$. In other words we can see a very clear core-periphery emerging, whereby cooperators form a core and defectors are typically isolated in the periphery. This mesoscale structure has been seen elsewhere in the literature, such as in multidimensional opinion spaces with only edge-breaking \cite{Shepherd2020}.

\begin{figure}
\centering
\includegraphics[width=119mm]{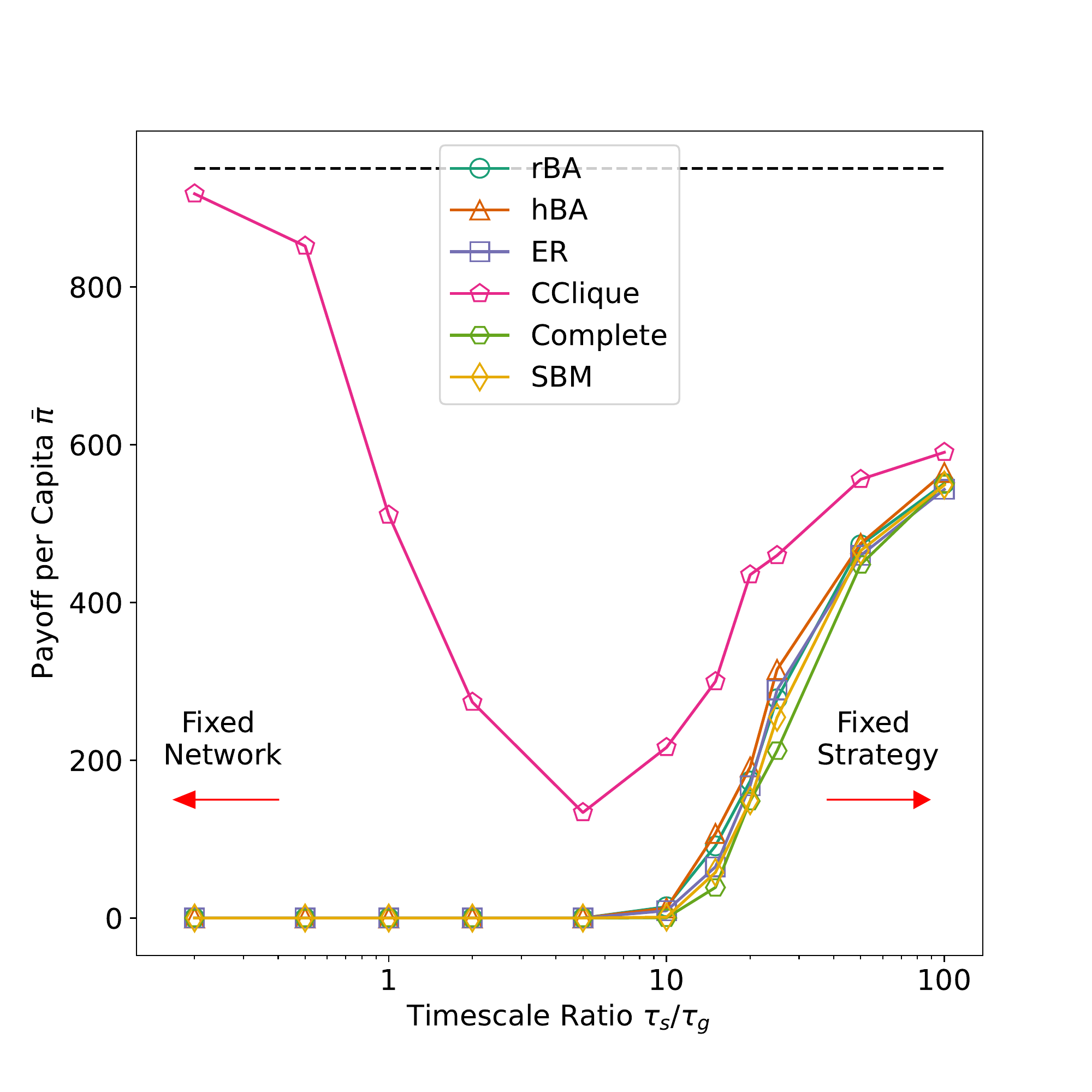}
\caption{Extreme popularity, imitate-payoff. After agents have played $5T$ rounds of a Prisoner's Dilemma, following GM1 and BM1, the final payoff per capita is plotted for different ratios of strategic to graph-theoretic timescales $\tau_s/\tau_g$. Each point is an ensemble average over $100$ simulations, with colour denoting the initial condition; the black dashed line is a baseline if all agents cooperated and were on a complete graph. Note towards the fixed strategy limit the rate of cooperation is bounded-from-above by the initial level of $C=15$}
\label{fig:timescale_ratio_g1b1}
\end{figure}

Once again we emphasise the emergence of a core-periphery, despite the difference in partner-update rules. In fact once strategies are allowed to change, as we will see in the next section, the two absorbing fixed points of the system are in effect core-peripheries of $C=0$ and $C=N$. That is, either the system turns into a complete graph of only cooperators, or an empty graph of only defectors; the stability of these two states is governed entirely by what strategy-update rules are implemented, i.e. the assumptions of how people decide to cooperate or defect. 

\subsection{Coevolutionary Process}
Here we address the cases when both edges and strategies can update, and do so on timescales of $\tau_g$ and $\tau_s$ respectively. By varying the ratio of timescales, $\tau_s/\tau_g$, we are able to explore the bifurcations that occur and regard the threshold effect that emerges. 

\begin{figure*}[tbhp]
\centering
\includegraphics[width=119mm]{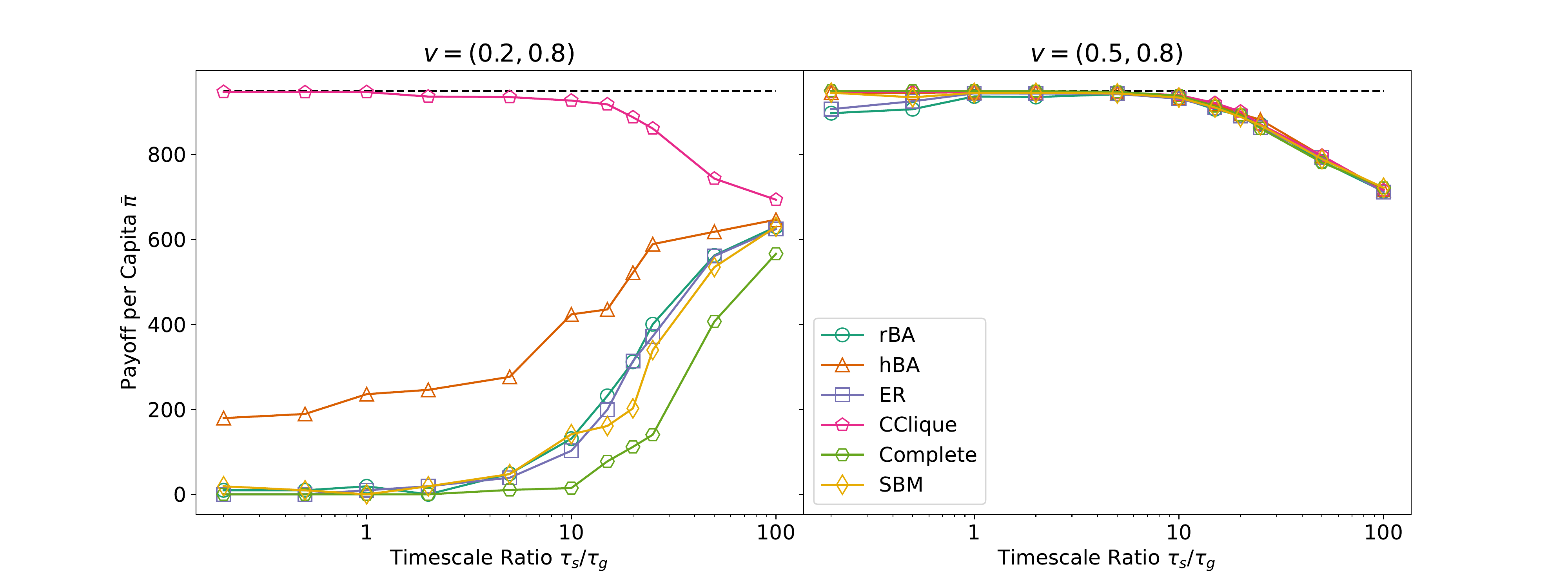}
\caption{Extreme popularity, conditional cooperation. After $5T$ rounds of a Prisoner's Dilemma, following GM1 and BM2, the final payoff per capita is plotted for different ratios of strategic to graph-theoretic timescales $\tau_s/\tau_g$. Each point is an ensemble average over $100$ simulations, with colour denoting the initial condition; the black dashed line is a baseline if all agents cooperated and were on a complete graph. Note towards the fixed strategy limit the rate of cooperation is bounded-from-above by the initial level of $C=15$. Cooperators find it easy to defect in the left panel, requiring only $20\%$ of neighbours are defectors. In the right panel a switch to defection requires a majority of neighbours to already be defectors}
\label{fig:timescale_ratio_g1b2}
\end{figure*}

In particular we focus on combinations of extreme popularity (GM1) rule with the three behavioural models to produce very different timescale curves (Fig. \ref{fig:timescale_ratio_g1b1}-\ref{fig:timescale_ratio_g1b3}), by simulating 100 simulations for each combination at different ratios of strategic to graph-theoretic timescales $\tau_s/\tau_g$. After $5T$ rounds the ensemble-average payoff per capita is measured and plotted. A commonality to note is towards the fixed strategy limit, the rate of cooperation tends towards the initial $C=15$ precisely because exceedingly few people have changed from cooperate to defect (or vice versa).

For comparison, in experiments where a fraction $k$ of subject pairs - such as in \cite{Wang2012} and \cite{Sohn2019} - are picked at random to update every round, $\tau_s/\tau_g = k$. In much of the experimental literature \cite{Rand2011,Rand2014,Sohn2019}, there are 3 typical values for $k$: the fixed ($k=0\%$), viscous ($k=10\%$) and fluid ($k=30\%$) conditions. They all found that the fluid condition has higher levels of cooperation than either the fixed or viscous cases, in other words higher $k$ have higher cooperative levels. We replicate these results in with simulations across the different update rules (see Figures \ref{fig:timescale_ratio_g1b1} and \ref{fig:timescale_ratio_g1b2}) below and show how other unseen phenomena may also occur.

\subsubsection*{Imitate-Payoff}
As seen in Fig. \ref{fig:timescale_ratio_g1b1}, there is a separation of time scales in most realistic graphs under extreme popularity (GM1) and imitate-payoff (BM1). After such a long period of time most of the realistic graph models behave near identically with mass defection occurring consistently for $\tau_s/\tau_g < 10$ and non-zero levels of cooperation subsiding above this threshold. The difference in behaviour between the static network limit and the dynamic network - that the fraction of cooperators is higher in the dynamic case - has been observed in experiments \cite{Wang2012}. 

However for the CClique initial graph - a highly artificial/pathological network where all cooperators begin as friends with all defectors entirely disconnected - playing under the same rules a reverse Goldilocks zone appears. That is for intermediate values of $\tau_s/\tau_g$ defection is rife, while towards the fixed network limit cooperation is nearly maximal.

From an individual's perspective placed in a realistic graph, defection is the optimal and preferential strategy - the population is mixed enough that defectors can infect cooperators. As such when strategies can update quickly enough, all players have enough time to defect hence mass defection occurs. However when strategies are slow to update, there simply is not enough time for everyone to defect; consider when $\tau_s/\tau_g = 10$, although all pairs have been updated around 5 times each, only around half of possible imitation updates have occurred.

In contrast, the CClique condition actually promotes cooperation at the individual level, at least in the early stages, precisely due to the core being resilient against defection. Note that the strategy vector $\bm{s}$ will only start to change once a defector has attached to the core - otherwise there will be no alternative strategy to imitate from. Provided the core is large enough, when strategies update rapidly the likelihood of a defector imitating a cooperator is far higher than the reverse, in other words the core converts defectors quicker than they can infiltrate the core. As the edges update quicker, more and more defectors can attach themselves to the core quickly enough to start converting the cooperators. After some point strategies become too slow for everyone to defect hence the payoff per capita rises again in the limit of fixed strategy. These two competing factors thus produce a reverse Goldilocks zone, where cooperation is minimised, not maximised, at intermediate ratios.

Finally note that for all realistic initial conditions when the network is static ($\tau_s/\tau_g=0$), we reproduce the qualitative result, seen empirically in \cite{Gracia-Lazaro2012a,Gracia-Lazaro2012b} and theoretically in \cite{Suri-Watts-2011}, that static networks do not promote cooperation. In fact this behaviour appears again for different strategy update rules, such as the conditional cooperation in the left of Fig. \ref{fig:timescale_ratio_g1b2}. However let us be clear that this is not always the case, across all static networks and across all update rules; we can qualitatively capture the discrepancies seen in the literature.

\subsubsection*{Conditional Cooperation}
As shown by Fig. \ref{fig:timescale_ratio_g1b2} the choice of behavioural model, which \textit{a priori} is as reasonable as any other, may produce vastly different results. For $v = (0.2,0.8)$, that is when cooperators easily defect, we see a similar curve for most realistic initial conditions as in the imitate-payoff case - heavy defection for low ratios with higher cooperation at higher ratios. However we already see differences, for a start when cooperators have the highest degree in a scale-free graph non-zero cooperative levels are maintained at all $\tau_s/\tau_g$. This directly contrasts the rBA condition, where cooperators are randomly assigned in a scale-free graph. This result alone implies possible policy implications in order to sustain global cooperation.

Second, unlike in Fig. \ref{fig:timescale_ratio_g1b1}, the CClique condition no longer produces a minimum in cooperators, instead there is a clear separation where mass cooperation occurs. If the simulations were to run longer then we would see mass cooperation at all ratios, whereas for the realistic initial graphs mass defection would occur. This suggests that for this parameter set, this rule combined with the network structure heavily promotes cooperation at all times.

Third, we see how the parameter values control the extent to which cooperation is promoted, looking at the right of Fig. \ref{fig:timescale_ratio_g1b2}. This happens because the threshold to defect is now much higher, so that cooperators remain cooperative, while defectors begin to cooperate, thus cooperative rate only ever increases.

\subsubsection*{Pairwise Comparison}
\begin{figure}[bthp]
\centering
\includegraphics[width=119mm]{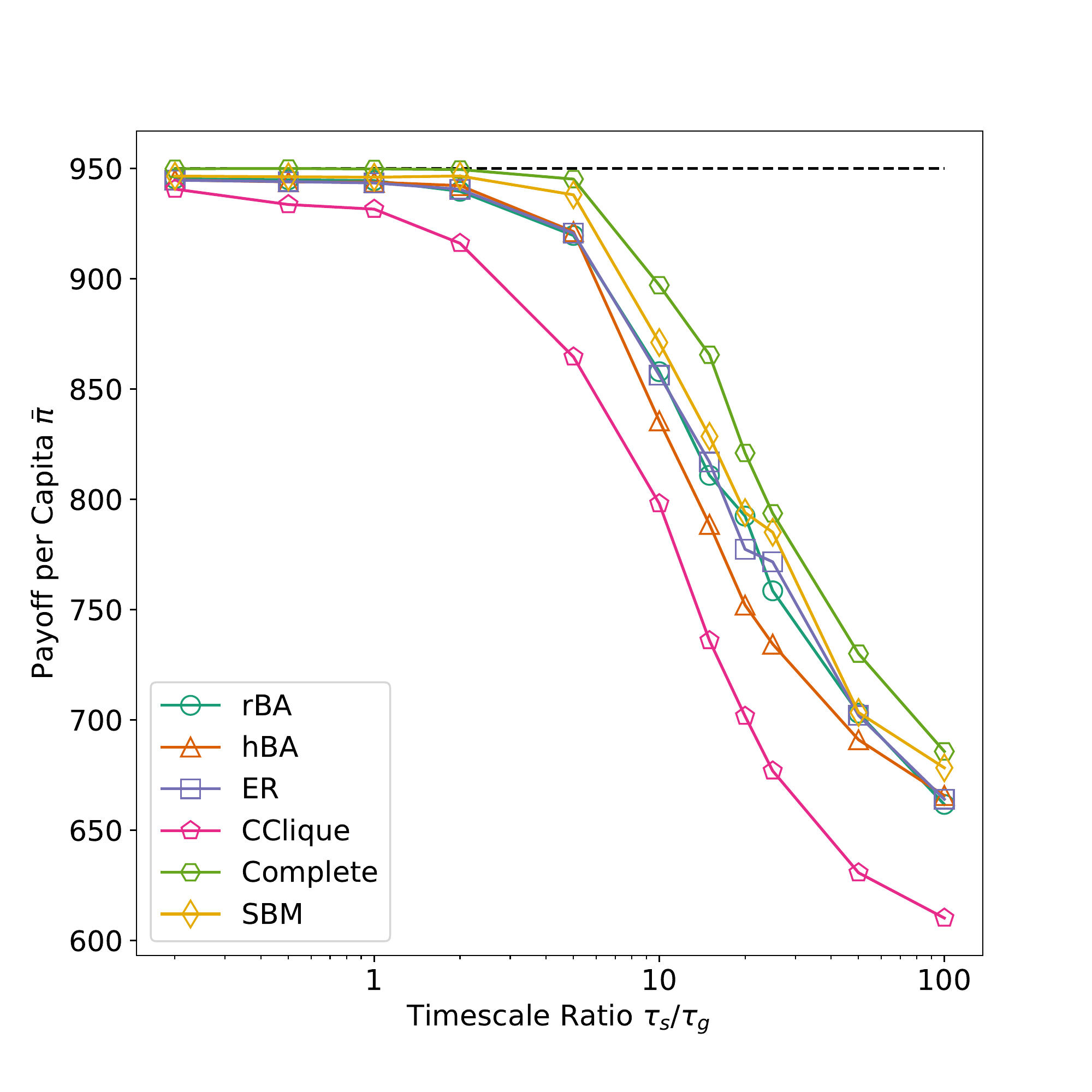}
\caption{Extreme popularity, pairwise comparison. After agents have played $5T$ rounds of a Prisoner's Dilemma, following GM1 and BM3, the final payoff per capita is plotted for different ratios of strategic to graph-theoretic timescales $\tau_s/\tau_g$. Each point is an ensemble average over $100$ simulations, with colour denoting the initial condition; the black dashed line is a baseline if all agents cooperated and were on a complete graph. Note towards the fixed strategy limit the rate of cooperation is bounded-from-above by the initial level of $C=15$}
\label{fig:timescale_ratio_g1b3}
\end{figure}

For $\beta=1$ the timescale curve of the pairwise comparison model, Fig. \ref{fig:timescale_ratio_g1b3}, behaves similarly to the conditional cooperation for high defecting threshold. That is cooperation is promoted across all $\tau_s/\tau_g$ likely as there are already many cooperators to begin with so that a defector feels an immense amount of peer pressure to cooperate. In this case when $\beta$ is lower, the resultant level of cooperation will be similarly smaller, as there is less pressure to imitate them; otherwise increasing $\beta$ would increase the speed at which mass cooperation happens. Moreover, we suspect the role of initial number of cooperators to be vital, that varying $C_0 \equiv C(0)$ will lead to a bifurcation in $C_\infty \equiv C(t\to\infty)$. Already two bifurcation points are trivial, $C_0 = 0$ and $C_0 = N$, as in both cases there are no alternative views to copy hence $C_\infty = C_0$.

Interestingly here, the CClique initial condition produces the lowest payoff per capita - unlike under different behavioural models. Even nearer the fixed strategy limit, the payoff per capita is around $50$ units short of the realistic networks - this suggests that inside the core roughly one cooperator has had the chance and decided to defect. 

Consider the CClique's evolution at intermediate times - that is when defectors have connected to the core but no agent has changed strategies - where there are $C_0$ cooperators and each cooperator has on average $d$ defecting friends (hence each defector has on average $C_0 d/D$ cooperative partners. For a cooperator $i$ and a defector $j$, the expected payoffs are $\pi_i = 50(C_0 - 1 -d)$ and $\pi_j = 100C_0d/(N-C_0)$, such that the probability for $i$ to imitate $j$ following Eq. \ref{eq:pairwise_comparison} and given the conditions of our experiment, is as below.
\begin{equation*}
    p_{ij} = \frac{1}{1+\exp{[-200(d-2)]}}
\end{equation*}

Here we see immediately that it becomes increasingly likely a cooperator $i$ will defect, once $d\geq 2$ defectors have connected to her. Moreover, when $\tau_s/\tau_g >> 1$, we can use Eq. \ref{eq:aij_extreme_popularity} to estimate $d(t) \approx (1-\exp(-t/\tau_g))D/2$ - so that $d$ grows rapidly past $d=2$ towards $d=D/2$. In other words cooperators will be under immense pressure to defect. However as $\tau_s$ is large, only very few agents have the opportunity to update their strategies, hence we typically only see 1 cooperator defect.

\section{Discussion}
By exploring different, but equally plausible, update rules - a proxy for hypotheses about agents' behaviour - we have observed the emergence of qualitatively contrasting phenomena surrounding cooperation, that have each been reported by various theoretical and experimental works. In using a singular framework, CANDY, we are able to isolate which assumptions and to what end they promote mass cooperation, suggesting that discrepancies in the literature arise from different assumptions and experimental design than being necessarily descriptive of real-world behaviour. 

Through our framework we have been able to reconstruct the qualitative results of many previous works, even where they may at first glance seem contradictory. For a start, multiple experiments \cite{Rand2011,Rand2014}, agent-based simulations \cite{Santos2005} and theory \cite{Ohtsuki2006,Pinheiro2016,Fotouhi2019} show that non-zero levels of cooperation are maintained when strategies are updated sufficiently quick ($\tau_s/\tau_g$ small) or similarly when networks are fully static; we see this across all initial networks for the pairwise comparison rule (Fig. \ref{fig:timescale_ratio_g1b3}) and for the conditional cooperation under the right parameter regime (right of Fig. \ref{fig:timescale_ratio_g1b2}). Moreover, for some rules we considered, we see a threshold effect in the timescales that has been theorised by \cite{Pacheco2006} and \cite{Santos-et-al-2006} while in other cases we see a distinct lack of one as in the conditional cooperation (right of Fig. \ref{fig:timescale_ratio_g1b2}) reported by \cite{Wang2012}. Finally we are able to replicate some of the empirical trends seen in experiments \cite{Rand2011,Wang2012}, that for higher $\tau_s/\tau_g$ a higher rate of cooperation is maintained.

Our analysis suggests that although the speed of interactions (both behavioural and relational) is an important factor, it is not in general a sufficient condition for mass cooperation as suggested in \cite{Li2020}, and is largely a by-product of the finite nature of the game. In the majority of conditions, we observed a threshold effect in the relative update speed $\tau_s/\tau_g$; however given infinite time for any ratio of timescales, such a 'threshold' disappears, and mass cooperation/defection depends upon the condition. For example as seen in the conditional cooperation with extreme popularity, the initial network structure matters immensely and can either promote cooperation or promote defection. In other words, time-permitting, what truly matters is the context, the game structure(s) and the decision-making style of agents, \textit{not} the relative speed of updates. 

The one exception is in the single case of a so-called Goldilocks zone in defectors, where, for the pathological CClique initial network, a near static network promotes cooperation while rapid edge updates seem to favour defection. In the infinite time limit we would expect to see a threshold emerging. 

Notable future research avenues include the study of games other than the Prisoner's Dilemma and the analysis of real-world group formation, such as working groups exhibiting hierarchical compartmentalisation and shifting pairwise interaction (e.g., Facebook or MS Teams).

Understanding the core-peripheral structures that emerge in real-world interaction, where cooperators collaborate closely with one another and defectors are ostracised, may have useful policy implications. 

A further avenue for future work would be to explore a protocol whereby defectors are slowly engaged and introduced into the cooperator-core, to reduce the temptation on a cooperator to defect and to ease the defector into a more positive mindset.




\section*{Appendix A - From Variables to Expectations}
As both $a_{ij}$ and $s_i$ are binary variables, we can instead consider them as independent\footnote{Even though the distribution parameters follow coupled differential equations, the random variables are importantly pointwise independent and thus statistically independent.} Bernoulli random variables, for all $i,j\in V$. In doing so, we can work in the continuous space of probabilities (specifically the parameter-space of Bernoulli distributions) which is equivalent to the expectations of our random variables, rather than the discrete space $\{0,1\}^{N(N+1)}$. By taking expectations, we can more easily analyse agent-based models by comparing the ensemble average with the statistical expectation.
\begin{gather}
    a_{ij} \sim \text{Bern}(\tilde{a}_{ij}) \\
    s_i \sim \text{Bern}(\tilde{s}_{i})
\end{gather}


As seen in Eqs. \ref{eq:C}-\ref{eq:pibar} we require distributions of Bernoulli variables, the sums, the products and the sum of products thereof. Fortunately by Proposition \ref{thm:Bernoulli_product}, assuming $a_{ij}$ and $s_j$ are independent, $a_{ij}s_j$ is Bernoulli. The sum of such a product are thus Poisson Binomial by definition; Proposition \ref{thm:PB_product} is a more general statement of this fact, the product of two Poisson binomial variables is itself Poisson binomial, in the terminology of Leemis \cite{Leemis2008} these distributions have the \textit{product} property. This is vital as we have an analytic form for the expectation of a Poisson binomial variable $Z\sim PB(p_1,\cdots,p_n)$ given by $\mathbb{E}(Z) = \sum_{i=1}^np_i$ \cite{PoissonBinomial}. Proofs of both propositions can be found in \cite{Leemis2008}.

\begin{proposition}
    Let $X$ and $Y$ be two Bernoulli random variables $X\sim \text{Bern}(p)$ and $Y\sim \text{Bern}(q)$ with success probabilities $p$ and $q$ respectively. The product $Z=XY$ is also Bernoulli $Z\sim \text{Bern}(pq)$ \cite{Leemis2008} with success probability $pq$. \label{thm:Bernoulli_product}
\end{proposition}

\begin{definition}[Poisson binomial distribution \cite{PoissonBinomial}]
Let $S_n$ be the sum of $n$ independent, possibly non-identical Bernoulli random variables $X_1,\cdots,X_n$ with success probabilities $p_1,\cdots,p_n$ respectively. $S_n$ is a Poisson binomial distribution, $S_n \sim \text{PB}(p_1,\cdots,p_n)$.
\end{definition}

\begin{proposition}
    The product of two Poisson binomial distributions is Poisson binomial. Let $Z_X$ and $Z_Y$ be two Poisson binomial random variables, formed from $n$ independent non-identical Bernoulli random variables $\{X_1,\cdots,X_n\}$ and $\{Y_1,\cdots,Y_n\}$ respectively, with success probabilities $\bm{p^X}=(p^X_i)^n_{i=1}$ and $\bm{p^Y}=(p^Y_i)^n_{i=1}$. The product $Z = Z_XZ_Y = \sum_{i,j} X_iY_j$ is Poisson binomial $Z\sim PB(\bm{p^X},\bm{p^Y})$. \label{thm:PB_product}
\end{proposition}

As $C$ is the sum of $N$ independent non-identical Bernoulli distributions it is Poisson binomial, $C\sim PB(\bm{\tilde{s}})$, with mean $\sum_{i\in V} \tilde{s}_i$ \cite{PoissonBinomial}. Moreover, by Proposition \ref{thm:PB_product}, the per capita payoff is similarly Poisson binomial, with mean $\sum_{i,j\in V}\tilde{a}_{ij}\tilde{s}_j(b-c)/N$. In other words we can replace the local variables on the right hand side of Eq. \ref{eq:C} and \ref{eq:pibar} with their respective probabilities to find the expectations of the global variables, which for brevity we also denote as $C$ and $\bar{\pi}$. 

\section*{Appendix B - Dynamical Systems} 
For a differential equation $\dot{x} = F(x,t)$, defined by the vector field $F$, on a space $X$ the solution is uniquely the flow $\Phi : X \times \mathbb{R} \rightarrow X$ that maps a point $x\in X$ to its position a time $t\in\mathbb{R}$ later.  

\begin{definition}[Flow]
A flow $\Phi$ on a space $X$ is a continuous mapping $\Phi:X\times \mathbb{R} \rightarrow X$ such that for all $x\in X$ and for all $s,t\in \mathbb{R}$:
\begin{gather*}
    \Phi(x,0) = 0 \\
    \Phi(\Phi(x,t),s) = \Phi(x,t+s)
\end{gather*}
\end{definition}

Let $\tilde{\mathcal{A}} = [0,1]^{N\times N}$ be the space of $N\times N$ expected adjacency matrices such that the $ij^{th}$-element is the probability an edge exists between $i$ and $j$. Equivalently one can think of this as the space of $N\times N$ adjacency matrices with bounded weights. Similarly let $\tilde{\mathcal{S}}=[0,1]^N$ be the space of $N$ probabilities to cooperate, or equivalently the space of $N$ continuous strategies. Conceptually we are thus considering a point $(\tilde{A},\bm{s})\in \tilde{\mathcal{A}}\times \tilde{\mathcal{S}}$ in the product space, that moves continuously due to the vector field $F$, which represents the coevolutionary process.

Analogously to flows and vector fields in $\mathbb{R}^2$ which have an $x$-component and a $y$-component, our velocity $F$ has an `adjacency-component', $f^g$, and a `strategy-component' $f^s$; each `component' is in fact either matrix-valued or vector-valued. Assuming the process does not explicitly depend on node-labels we can further assign two a time-scale to each component: a graph-theoretic $\tau_g$ and a strategic $\tau_s$.








\bibliographystyle{spmpsci}
\bibliography{references}

\end{document}